\documentclass[12pt,letterpaper,english]{article}
\usepackage[T1]{fontenc}
\usepackage[latin1]{inputenc}
\usepackage{color}
\usepackage{graphicx}
\usepackage{amssymb}

\makeatletter

\usepackage{osajnl2}
\usepackage[draft]{hyperref}
\usepackage{babel}
\makeatother

\begin{document}

\title{Ultra-Broadband Coherence-Domain Imaging Using Parametric Downconversion
and Superconducting Single-Photon Detectors at 1064 nm}

\maketitle

\author{Nishant Mohan,$^{1,*}$ Olga Minaeva,$^{2,3}$ Gregory N. Goltsman,$^{3}$
Mohammed F. Saleh,$^{2}$ Magued B. Nasr,$^{2}$ Alexander V.
Sergienko,$^{2,4}$ Bahaa E. A. Saleh,$^{2,5}$ and Malvin C.
Teich$^{1,2,4}$ }

\address{$^1$Department of Biomedical Engineering,\\ Boston University, Boston MA 02215}
\address{$^2$Department of Electrical and Computer Engineering,\\ Boston University, Boston MA 02215}
\address{$^3$Department of Physics, \\Moscow State Pedagogical University, Moscow 119992, Russia}
\address{$^4$Department of Physics,\\ Boston University, Boston MA 02215}
\address{$^5$College of Optics and Photonics (CREOL),\\ University of Central Florida, Orlando FL 32816}
\address{$^*$Corresponding author: nm82@bu.edu}
\begin{abstract}
Coherence-domain imaging systems can be operated in a single-photon
counting mode, offering low detector noise; this in turn leads to
increased sensitivity for weak light sources and weakly reflecting
samples. We have demonstrated that excellent axial resolution can be
obtained in a photon-counting coherence-domain imaging (CDI) system
that uses light generated via spontaneous parametric down-conversion
(SPDC) in a chirped periodically poled stoichiometric lithium
tantalate (chirped-PPSLT) structure, in conjunction with a niobium
nitride superconducting single-photon detector (SSPD). The bandwidth
of the light generated via SPDC, as well as the bandwidth over which
the SSPD is sensitive, can extend over a wavelength region that
stretches from 700 to 1500 nm. This ultra-broad wavelength band
offers a near-ideal combination of deep penetration and ultra-high
axial resolution for the imaging of biological tissue. The
generation of SPDC light of adjustable bandwidth in the vicinity of
1064 nm, via the use of chirped-PPSLT structures, had not been
previously achieved. To demonstrate the usefulness of this
technique, we have constructed images for a hierarchy of samples of
increasing complexity: a mirror, a nitrocellulose membrane, and a
biological sample comprising onion-skin cells.
\end{abstract}


\section{Introduction}

Over the past decade, coherence-domain imaging (CDI), particularly
optical coherence tomography (OCT), has come into its own. OCT
relies on the interference of partially coherent light to achieve
axial sectioning \cite{salehteichbook}. The technique has been
highly successful, particularly in biology and medicine
\cite{brezinski2006,fercher2003,bopart1998}. Indeed, in recent
years, OCT has joined the armamentarium of diagnostic imaging
techniques, and has found use in clinical ophthalmology
\cite{HeeOptho95}, dermatology \cite{WelzelDerm01}, cardiology
\cite{TearneyCardio03}, and gastroenterology \cite{BoumaGI06}. In
biological tissue, OCT typically achieves an axial resolution of a
few $\mu$m and achieves imaging to depths of 2--3~mm
\cite{Drexler04}.

The central wavelength at which a biological CDI system operates is
an important parameter; optical scattering in biological tissue
makes it difficult to image deeply in the visible region so that
most CDI systems make use of light sources with wavelengths longer
than 700 nm. The limitation on the long-wavelength side is governed
by the absorption of water, which becomes substantial at about 1500
nm. Moreover, since the axial resolution of a CDI system improves as
the spectral bandwidth of the light source increases, use of the
entire wavelength range from 700 to 1500 nm provides a suitable
combination of deep penetration and ultra-high resolution for
biological tissue \cite{Mohan08}.

In this paper, we demonstrate that coherence-domain imaging may be
achieved at a low photon-flux level by making use of light generated
by spontaneous parametric down-conversion (SPDC) in a chirped
periodically poled stoichiometric lithium tantalate (chirped-PPSLT)
structure, via phase matching to the fundamental. This structure is
pumped with 532-nm CW light from a high-power laser to generate
broadband light centered at 1064 nm. Broadband SPDC has been
generated previously in chirped periodically poled structures, but
only in the 800-nm region and only via phase matching to the third
harmonic, which is less efficient
\cite{Carrasco06,Harris07,Nasr08PRL,Nasr08OE}. We have built on our
previous efforts \cite{Mohan08} in carrying out photon-counting
optical coherence-domain reflectometry with superconducting
single-photon detectors (SSPDs), which are sensitive over the entire
spectral range of interest for biological imaging, namely 700--1500
nm.

Section~II summarizes the theory of the SPDC optical source, along
with optical spectra calculated for parameter ranges of interest.
This is followed in Sec.~III by the experimental spectra obtained
from the chirped structure, as well as the results of
photon-counting CDI experiments that make use of this source. In
particular, we display CDI interferograms for a hierarchy of samples
of increasing complexity: a mirror, a pellicle, and a biological
sample comprising onion-skin tissue. In the discussion provided in
Sec.~IV, we compare the relative merits of using the chirped-PPSLT
SPDC source and SSPD detector for CDI with other more conventional
sources and detectors.

\section{Design of Downconversion Source}\label{sec:simspect}

Consider the process of SPDC in which a pump wave at frequency
$\omega_{p}$ is down converted to signal and idler waves at
frequencies $\omega_{s}$ and $\omega_{i}$, respectively, such that
$\omega_{p}=\omega_{s} + \omega_{i}$. The power spectral density
$S(\omega_{s})$ of the downconverted light (signal or idler) is
governed by the phase-matching condition in the nonlinear structure,
which, for collinear emission is given by \cite{salehteichbook}
\begin{equation}
S(\omega_{s})\propto\left|\int_{0}^{L}d(z)\,\exp[-j\,\Delta
k(\omega_{s})\,z]\, dz\right|^{\,2},\label{eq:Phase Matching}
\end{equation}
where $z$ is the distance along the direction of propagation of the
waves, $L$ is the overall length of the structure, $d$ is the
effective second-order nonlinear coefficient of the material, and
$\Delta k$ is the phase mismatch between the pump and the
downconverted waves.

Both the local period of the nonlinear coefficient $d(z)$ and
$\Delta k$ depend on temperature $T$, so that the power spectral
density is also a function of temperature. The dependence of $\Delta
k$ on the signal frequency and temperature is provided by
\begin{equation}
\Delta
k(\omega_{s})=\frac{n(\omega_{p}\,,T)\omega_{p}}{c}-\frac{n(\omega_{s}\,,T)\omega_{s}}{c}
-\frac{n(\omega_{p}-\omega_{s}\,,T)(\omega_{p}-\omega_{s})}{c}\,,
\end{equation}
where $c$ is the speed of light in vacuum and $n$ is the refractive
index of the material; its dependence on frequency and temperature
can be determined from the temperature-dependent Sellemeier equation
\cite{SelEq}.

The function $d(z)$ in Eq.~(\ref{eq:Phase Matching}) for a structure
with chirped poling is given by
\begin{equation}
d(z)=\sum_{k=1}^{N}s(z-a_{k},b_{k}),\label{eq:d(z)}
\end{equation}
where $N$ is the total number of periods. The quantity $b_{k}$ in
Eq.~(\ref{eq:d(z)}) represents the width of the $k$th period. For a
bipolar rectangular pulse, $s(z,b)$ is given by
\begin{equation}
s(z,b)=\left\{ \begin{array}{cc}
1, & 0<z\leq b/2\\
-1, & b/2<z\leq b\\
0, & \mathrm{{\displaystyle {\textstyle otherwise.}}}\end{array}\right.
\end{equation}
For a crystal poled with a linearly chirped spatial frequency,
$b_{k}$ is given by $1/b_{k}=1/b_{1}-(k-1)\zeta$, where $\zeta$ is
the chirp parameter. The term $a_{k}$ in Eq. (\ref{eq:d(z)})
represents the starting location of the $k$th period and is
represented by the iterative relation $a_{k}=a_{k-1}+b_{k-1}$ with
$a_{1}=0$. The features of the structure, i.e., $b_{k}$ for $1<k\leq
N$, is defined entirely by the length of the first period $b_{1}$
and the chirp parameter $\zeta$.

Our calculations were carried out for stoichiometric lithium
tantalate (SLT) as the nonlinear optical material. The parameters
used for the calculations were the same as those of the actual
structures used in our experiments, which we report in
Sec.~\ref{ssec:obsspect}. The actual poling was conducted in such a
way that three separate structures were fabricated on the same
substrate, enabling us to carry out measurements for structures with
different chirp parameters; this also enabled us to better judge the
validity of our theory. The three structures have just about the
same overall length ($L\approx2$ cm) and number of periods
($N\approx2515$), but different chirp parameters $\zeta$ and lengths
of the first period $b_{1}$. The first of the three structures has
$b_{1}=7.95\,\mu\mathrm{m}$ and $\zeta=0$, i.e., it is unchirped.
The second has $b_{1}=7.85\,\mu\mathrm{m}$ and
$\zeta=1.26\times10^{-6}\:\mu\mathrm{m}^{-1}$, whereas the third
structure has the highest chirp with $b_{1}=7.5\,\mu\mathrm{m}$ and
$\zeta=6.24\times10^{-6}\:\mu\mathrm{m}^{-1}$. For SLT, the
temperature dependence of the width of each period can be
accommodated via the relation:
$b_{k}(T)=b_{k}(25)\,[\alpha(T-25)+\beta(T-25)^{2}]$, where $\alpha$
and $\beta$ are thermal expansion coefficients and $T$ is the
temperature in degrees Celsius \cite{SelEq}.

The parameters $b_{1}$ and $\zeta$ enable us to determine $d(z)$
while the temperature-dependent Sellmeier equation, along with the
pump frequency $\omega_{p}=5.6\times10^{14}$ Hz (corresponding to a
pump wavelength of 532 nm), provides us with $\Delta k$. Inserting
$d(z)$ and $\Delta k$ in Eq.~(\ref{eq:Phase Matching}) enables us to
obtain the estimated power spectral density of the output signal
beam for different temperatures, via numerical evaluation. The
wavelength ranges of spectral components that are present, at
various temperatures, are displayed as brightness images in Fig.~1
for the unchirped structure [Fig. 1(a)], the medium-chirped
structure [Fig. 1(b)], and the maximum-chirped structure [Fig.
1(c)]. The calculated normalized spectrum is displayed in Fig. 1(d)
for the maximum-chirped structure at a temperature of 80$^\circ$ C.

\section{Experiments}

\subsection{Generic experimental arrangement} \label{ssec:genexp}

The generic interferometric arrangement for the experiments we
report in this Section is schematized in Fig.~2(a). The experiments
were carried out using either single-mode-fiber-coupled
downconversion or single-mode-fiber-coupled light from a
superluminescent diode (SLD). The broadband light emanating from the
source was coupled into a single-mode (SM) fiber and collimated by
lens L3 before being fed into the Michelson interferometer. The
light in the interferometer was focused onto the reference mirror
and the sample using lenses L4 and L5, respectively. Lens L4 and the
reference mirror were placed on a single nanopositioning stage while
the sample was placed on another one. Their positions were changed
in accordance with the arrows shown in the figure. The light exiting
the interferometer was fed into either a single-photon avalanche
detector (SPAD) or a SSPD via fiber-coupling and a lens (not shown).
It is worth noting that the optical components should be
transmissive over a broad band of wavelengths when working with
broadband optical imaging systems.

The downconversion source, shown in Fig.~2(b), consisted of light
from a cw frequency-doubled Nd$^{3+}$:YVO$_4$ laser (Coherent
Verdi), operating at a wavelength of 532 nm and at a power of 2 W,
that pumped a chirped-PPSLT device with three separate structures
fabricated on the same substrate, as described previously. The
structure was aligned to obtain collinear SPDC. The downconverted
light was collimated using a lens (L1) and coupled into a SM fiber
via another lens (L2). The filter, comprising a dichroic mirror and
a longpass filter, served to remove the pump and to allow only the
downconverted light to be coupled into the fiber. The light from the
SLD (Superlum SLD-47-MP) was centered at a wavelength of 930 nm and
had a full-width at half-maximum (FWHM) spectral width of 70 nm.

Optical imaging systems often make use of commercially available Si
or InGaAs semiconductor detectors, depending on the wavelength at
which the system operates. Roughly speaking, Si photodiodes are used
for wavelengths shorter than 1100 nm and operate best in the
vicinity of 800 nm, whereas InGaAs photodiodes, which are designed
for operation in the vicinity of 1300 nm, are used for wavelengths
longer than 1100 nm. The Si SPADs used in our experiments are
incorporated in commercially available single-photon counting
modules (Perkin-Elmer, Model SPCM-AQR-15-FC).

Inasmuch as neither Si nor InGaAs are sensitive over the entire
spectral range of 700--1500 nm, however, we have followed our
previous work and assayed the use of SSPDs in photon-counting
optical coherence-domain imaging \cite{Mohan08}. Such detectors
offer some benefits for single-photon-counting applications,
especially in the infrared, since they exhibit low dark-count rates
and operate in a spectral region that stretches from 0.4 to 6 $\mu$m
\cite{SSPDQE01}, which includes the region of biological interest.
The SSPDs used were fabricated from 4-nm-thick superconducting
niobium nitride (NbN) films \cite{SSPDFab03}; their operation has
been described elsewhere \cite{Mohan08,SSPDFast05}. The quantum
efficiency monotonically decreases with increasing wavelength of the
incident light. Measurements using SSPDs were carried out in the
same way as those described previously \cite{Mohan08}.

\subsection{Observed downconversion spectra for different chirp parameters}
\label{ssec:obsspect}

In this set of experiments, we measured the spectra associated with
the downconverted light emitted by the three different chirped-PPSLT
structures, as a function of temperature. The parameters for these
structures were provided in Sec.~\ref{sec:simspect}, where the
calculated spectra were reported (see Fig.~1). The experiments were
carried out using the setup displayed in Fig.~2, which makes use of
SM-fiber-coupled downconversion as the source, a SSPD as the
detector, and a mirror as the sample. The counts from the SSPD were
measured in a fixed time window, as a function of the reference-arm
displacement (position of the reference mirror). The resultant
interferograms were Fourier-transformed to obtain an estimate of the
power spectral density of the downconverted light from the
structures at different temperatures.

The experimental results are displayed as brightness images for the
unchirped, medium-chirped, and maximum-chirped structures in
Figs.~3(a), (b), and (c), respectively. The observed normalized
spectrum is displayed in Fig. 3(d) for the maximum-chirped structure
at a temperature of 80$^\circ$ C.

The spectral content of downconversion from the three chirped-PPSLT
structures, which are stable and predictable, roughly follow the
calculated patterns (see Fig.~1). For the unchirped structure, the
phase matching is satisfied for two narrow bands of wavelengths. The
widths of these bands increase for the medium-chirped structure, and
a broad wavelength band emerges for the maximum-chirped structure.
The spectra obtained for the unchirped and medium-chirped structures
are slightly broader than those predicted by the calculations. We
attribute this to the presence of the collimating lens (L1), which
collects the downconverted beams over a finite angular range, rather
than collinearly. Also, the measured optical power at longer
wavelengths is smaller than that expected on the basis of the
calculations. We believe that this arises because of a reduction in
the quantum efficiency of the SSPD, as well as changes in the
transmittance of the optical components, as the wavelength
increases.

These results illustrate that we have a sound understanding of the
features of SPDC light generation via chirped-PPSLT structures,
thereby allowing us to engineer broadband spectra as desired.

\subsection{Axial resolution for coherence-domain imaging
with different sources and detectors}
\label{ssec:res}

We next carried out a set of experiments to determine the relative
merits of using different sources of light and different detectors
for coherence-domain imaging.

After examining the axial resolution achievable by using broadband
SPDC (from the structure with the greatest chirp at a temperature of
80$^\circ$ C; see Sec.~\ref{ssec:obsspect}) in conjunction with a
SSPD, we demonstrate that the resolution is diminished when the SSPD
is replaced by a SPAD. We then proceed to show that the resolution
is further diminished when the downconversion source is replaced by
a SLD.

The axial resolution is determined by examining the widths of the
interferogram envelopes using a mirror as the sample, as well as by
making use of a pellicle as the sample.

In the first set of experiments to measure the axial resolution, we
used a mirror as the sample in the experimental arrangement depicted
in Fig. 2. We constructed interferograms by measuring the counts
from the detector in a fixed time window, as a function of the
position of the reference mirror (see Sec.~\ref{ssec:obsspect}),
using the three source/detector combinations indicated above. The
results are illustrated in Fig.~4. They reveal that the FWHM of the
interferogram envelopes increase from 1.6 to 2.8 to 6.3 $\mu$m for
the SPDC/SSPD, SPDC/SPAD, and SLD/SSPD combinations, respectively.
Since the narrower the width of the envelope, the better the
resolution, this demonstrates that the downconverted light with a
superconducting detector yielded the highest resolution of the three
source/detector combinations considered.

Another way of measuring the resolving power of an optical system is
to test its ability to distinguish reflections from two nearby
surfaces. The second set of experiments was similar to the first
set, as described above, except that we replaced the mirror by a
pellicle in the experimental arrangement depicted in Fig. 2. The
pellicle was a nitrocellulose membrane (Thorlabs) of refractive
index $n=1.5$ and thickness $L=2\:\mu$m.

Again we constructed interferograms using the three source/detector
combinations indicated above. The results are illustrated in Fig.~5.
It is evident that only the combination of a broadband
downconversion source and a superconducting detector (Fig.~5(a)) was
capable of clearly resolving the two surfaces of the membrane,
indicating that this combination offered the highest resolution. The
measured optical path length between the reflections is seen to be 3
$\mu$m in Fig.~5(a); this corresponds to a physical distance of $3/n
= 2\:\mu$m, which nicely matches the thickness of the pellicle.

In terms of resolution, then, these experiments lead us to conclude
that, in the vicinity of 1064 nm and of the source/detector
combinations we examined, the optimal combination for
coherence-domain imaging comprises broadband downconverted light
with a superconducting detector.

\subsection{Coherence-domain imaging of a biological sample}

Finally, we demonstrated the photon-counting coherence-domain
imaging of a biological sample, comparing the performance of two
source/detector combinations.

The sample was an onion-skin tissue from a white onion that was
adhered to a cover slip using a small drop of water. The sample was
placed in the generic experimental setup depicted in Fig.~2 and a
collection of A-scans were obtained at different transverse ($x$)
positions along the sample. At a given transverse position, the
lens--mirror combination (L4 and reference mirror) was scanned in
the $z$ direction over a range of 70 $\mu\mathrm{m}$, using a step
size of 100 nm, and counts were recorded at each step for a 500-ms
accumulation time. The duration of an A-scan was thus 350 sec.

The sample was then moved to the next transverse position, with a
step size of 5 $\mu\mathrm{m}$ in the $x$ direction, and the next
A-scan was recorded. The step size was chosen to be 5
$\mu\mathrm{m}$ since the transverse resolution of the imaging
system was estimated to be $\approx10\:\mu\mathrm{m}$, based on a
collimated beam width of $\thickapprox2.5\:\mathrm{mm}$, a focal
length of $\thickapprox25\:\mathrm{mm}$ for lens L5, and the central
wavelength of $1064\:\mathrm{nm}$. The sample was scanned in the $x$
direction over a range of 800 $\mu\mathrm{m}$, so that 160 scans
were collected in the axial direction. The limited photon flux
available from the downconversion source and the limited quantum
efficiency (5--12\%) of the particular SSPDs that we used in these
experiments \cite{Mohan08} result in long acquisition times. These
can be decreased, however, by making use of a more powerful laser
pump, a chirped SPDC structure with higher conversion efficiency,
optics that is optimized over a broad wavelength band, and a
detector with higher quantum efficiency.

The collected A-scans were used to construct an $xz$ B-scan of the
onion-skin sample using broadband SPDC (from the structure with the
greatest chirp at a temperature of 80$^\circ$ C; see
Sec.~\ref{ssec:obsspect}) in conjunction with a SSPD. The results
are illustrated in Fig.~6(a). The two surfaces of several onion-skin
cells are apparent (the size of an individual cell falls in the
range of onion-skin cells observed using other imaging modalities;
see, for example, Figs.~2(c), 4, and 6 of Ref.~\cite{Nasr09OC}). We
also constructed a B-scan of the same sample with the SPDC source
replaced by a standard SLD (see Sec.~\ref{ssec:obsspect}), again
using the SSPD. The results are displayed in Fig.~6(b). It is
apparent that the resolution is diminished when the SPDC source is
replaced by a SLD, by virtue of the narrower bandwidth of the
latter.

These results confirm once again that, in the vicinity of 1064 nm,
and of the source/detector combinations examined, the best
resolution was obtained by using broadband downconverted light and a
superconducting detector.

\section{Discussion}

Coherence-domain imaging systems can be operated in a single-photon
counting mode, offering low detector noise; this in turn leads to
increased sensitivity for weak light sources and weakly reflecting
samples \cite{Mohan08}. In this paper, we have demonstrated that
excellent axial resolution can be obtained in a photon-counting
coherence-domain imaging system that uses light generated via
spontaneous parametric down-conversion (SPDC) in a chirped
periodically poled stoichiometric lithium tantalate (chirped-PPSLT)
structure, in conjunction with a niobium nitride superconducting
single-photon detector (SSPD).

The bandwidth of the light generated via SPDC, as well as the
bandwidth over which the SSPD is sensitive, can extend over a
wavelength region that stretches from 700 to 1500 nm. This
ultra-broad wavelength band offers a near-ideal combination of deep
penetration and ultra-high axial resolution for the imaging of
biological tissue.

The generation of SPDC light of adjustable bandwidth in the vicinity
of 1064 nm, via the use of chirped-PPSLT structures, has not been
achieved previously. Unlike many other broadband sources, the
spectra are stable and predictable. We also showed that the observed
spectral characteristics of this source accord well with those
calculated based on a second-order nonlinear-optics model,
indicating that we have the capability of engineering broadband
near-infrared sources as desired.

There are a number of limitations associated with this system,
however: (1) the SPDC source is more complex and expensive than
commonly used SLDs; (2) SPDC generates substantially lower optical
photon flux than other commonly used broadband sources, such as
SLDs, fsec lasers, fiber lasers \cite{Lim02}, and supercontinuum
sources generated by photonic-crystal fibers \cite{Povazay07},
resulting in long data-acquisition times; and (3) the use of SSPDs
is complicated by their small active areas and low quantum
efficiencies, as well as by the need for cryogenic operation.

Nevertheless, in the vicinity of 1064 nm, we have explicitly
demonstrated that the axial resolution offered by the SPDC/SSPD
source/detector combination exceeds that achievable with either a
superluminescent diode (SLD) as the source, or a single-photon
avalanche detector (SPAD) as the detector. These findings were
confirmed by the coherence-domain imaging of a hierarchy of samples
of increasing complexity: a mirror, a nitrocellulose membrane, and a
biological sample comprising onion-skin cells.

\section{Acknowledgments}

This work was supported by the Bernard M. Gordon Center for
Subsurface Sensing and Imaging Systems (CenSSIS), an NSF Engineering
Research Center, and by a U.S.~Army Research Office (ARO)
Multidisciplinary University Research Initiative (MURI) Grant.

\clearpage{}

\section*{List of Figure Captions}

\noindent Fig. 1. Brightness images, in which brightness is
proportional to the calculated power spectral density of the
emission, at various temperatures. The features of the structures
are specified by the length of the first period $b_{1}$ and the
chirp parameter $\zeta$. (a) Unchirped structure with
$b_{1}=7.95\,\mu\mathrm{m}$ and $\zeta=0$. (b) Medium-chirped
structure with $b_{1}=7.85\,\mu\mathrm{m}$ and
$\zeta=1.26\times10^{-6}\:\mu\mathrm{m}^{-1}$. (c) Maximum-chirped
structure with $b_{1}=7.5\,\mu\mathrm{m}$ and
$\zeta=6.24\times10^{-6}\:\mu\mathrm{m}^{-1}$. (d) Calculated
normalized spectrum for the maximum-chirped structure at a
temperature of 80$^\circ$ C. The parameters were chosen to match
those of the structures used in the experiments described
subsequently. The bandwidth increases substantially with the chirp parameter.\\

\noindent Fig. 2. (a) The generic experimental arrangement makes use
of a Michelson interferometer comprising a beam-splitter (BS),
reference mirror, and sample. The broadband light emanating from the
source is coupled into a SM fiber and collimated by lens L3. The
light within the interferometer is focused onto the reference mirror
and the sample using lenses L4 and L5, respectively. The lenses,
reference mirror, and sample are placed on nanopositioning stages to
change their positions, as indicated by the arrows. Experiments were
performed using both SSPDs and SPADs as detectors. (b) The
downconversion source consists of light from a cw frequency-doubled
Nd$^{3+}$:YVO$_4$ laser (Coherent Verdi), operating at a wavelength
of 532 nm and at a power of 2 W, that pumps a chirped-PPSLT device.
The structure is aligned to obtain collinear SPDC. The downconverted
light is collimated using lens L1 and coupled into a SM fiber via
lens L2. The filter removes the pump light and allows only the
downconverted light to be coupled
into the fiber.\\

\noindent Fig. 3. Brightness images, in which brightness is
proportional to the measured power spectral density of the emission,
at various temperatures, from: (a) the unchirped structure; (b) the
medium-chirped structure; and (c) the maximum-chirped structure. (d)
Measured normalized spectrum for light from the maximum-chirped
structure at a temperature of 80$^\circ$ C. The results bear
considerable resemblance to the calculations displayed in Fig. 1.\\

\noindent Fig. 4. Normalized interferograms and their envelopes vs.
reference-arm displacement for a mirror sample (A-scans). In all
cases, the step size used in constructing the interferograms was 100
nm and the duration of the counting-time windows was 300 ms. (a)
Downconversion/superconducting-detector (SPDC/SSPD). The FWHM of the
interferogram envelope was 1.6 $\mu$m. The highest resolution was
achieved with this combination. (b)
Downconversion/avalanche-detector (SPDC/SPAD). The FWHM of the
interferogram envelope was 2.8 $\mu$m. (c)
Superluminescence/superconducting-detector (SLD/SSPD). The FWHM of
the interferogram envelope was 6.3 $\mu$m.\\

\noindent Fig. 5. Normalized interferograms vs. reference-arm
displacement for a pellicle sample (A-scans). In all cases, the step
size used in constructing the interferograms was 100 nm and the
duration of the counting-time windows was 300 ms. (a)
Downconversion/superconducting-detector (SPDC/SSPD). This
combination permitted reflections from the two surfaces to be
resolved. (b) Downconversion/avalanche-detector (SPDC/SPAD). The two
surfaces were not resolved. (c)
Superluminescence/superconducting-detector
(SLD/SSPD). The two surfaces were not resolved.\\

\noindent Fig. 6. Two-dimensional ($xz$) B-scans of an onion-skin
sample. (a) Scan collected using broadband downconversion light and
a superconducting detector (SPDC/SSPD). (b) Scan collected from the
same onion-skin sample using superluminescence light and the same
superconducting detector (SLD/SSPD). Higher axial resolution is
attained by using downconversion, by virtue of its broader
bandwidth. These cross-sectional views of the tissue highlight the
relatively large reflectances at cellular surfaces, which stem from
refractive-index discontinuities.\\

\clearpage{}
%
\begin{figure}[htbp]
 \centering \includegraphics[width=6in]{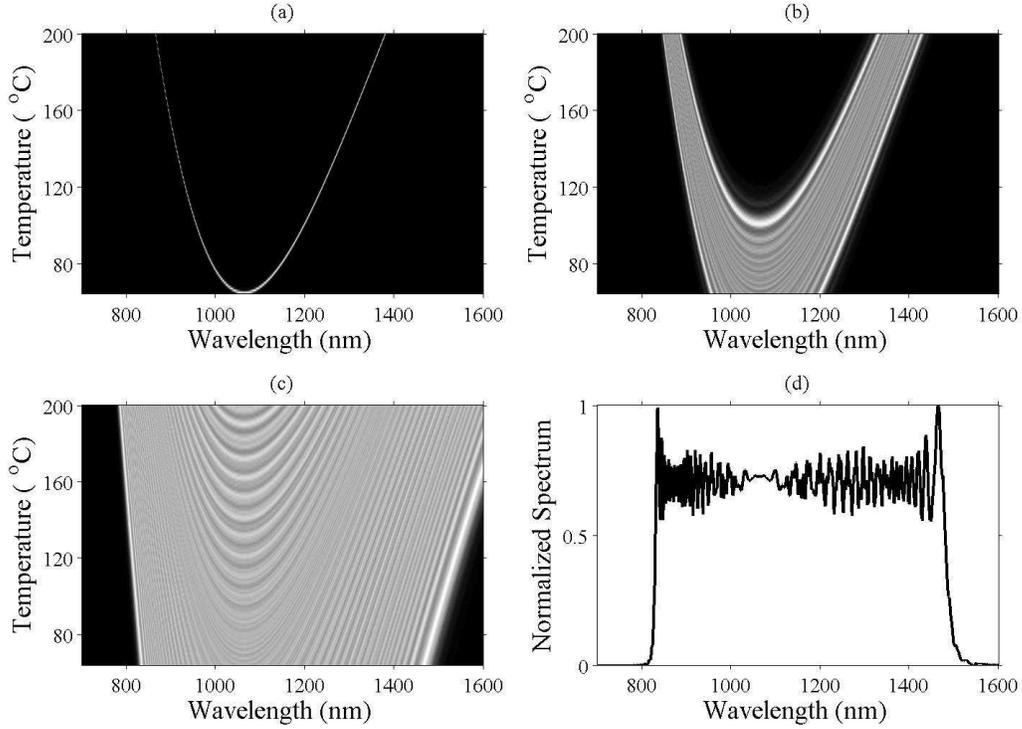}
\caption{Brightness images, in which brightness is proportional to
the calculated power spectral density of the emission, at various
temperatures. The features of the structures are specified by the
length of the first period $b_{1}$ and the chirp parameter $\zeta$.
(a) Unchirped structure with $b_{1}=7.95\,\mu\mathrm{m}$ and
$\zeta=0$. (b) Medium-chirped structure with
$b_{1}=7.85\,\mu\mathrm{m}$ and
$\zeta=1.26\times10^{-6}\:\mu\mathrm{m}^{-1}$. (c) Maximum-chirped
structure with $b_{1}=7.5\,\mu\mathrm{m}$ and
$\zeta=6.24\times10^{-6}\:\mu\mathrm{m}^{-1}$. (d) Calculated
normalized spectrum for the maximum-chirped structure at a
temperature of 80$^\circ$ C. The parameters were chosen to match
those of the structures used in the experiments described
subsequently. The bandwidth increases substantially with the chirp
parameter. Simulations-f1.eps.}
\end{figure}

\begin{figure}[htbp]
 \centering \includegraphics[width=6in]{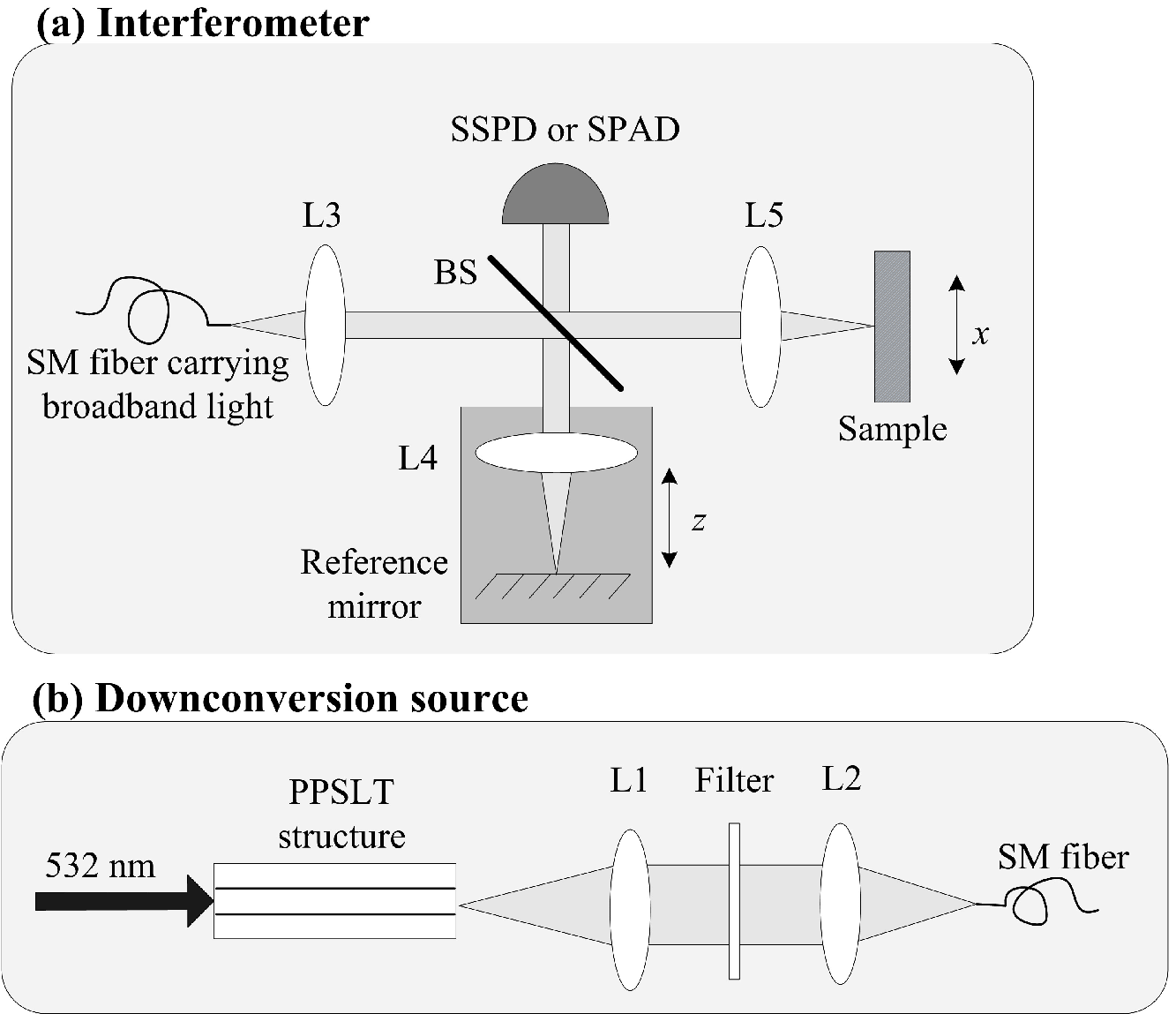}
\caption{(a) The generic experimental arrangement makes use of a
Michelson interferometer comprising a beam-splitter (BS), reference
mirror, and sample. The broadband light emanating from the source is
coupled into a SM fiber and collimated by lens L3. The light within
the interferometer is focused onto the reference mirror and the
sample using lenses L4 and L5, respectively. The lenses, reference
mirror, and sample are placed on nanopositioning stages to change
their positions, as indicated by the arrows. Experiments were
performed using both SSPDs and SPADs as detectors. (b) The
downconversion source consists of light from a cw frequency-doubled
Nd$^{3+}$:YVO$_4$ laser (Coherent Verdi), operating at a wavelength
of 532 nm and at a power of 2 W, that pumps a chirped-PPSLT device.
The structure is aligned to obtain collinear SPDC. The downconverted
light is collimated using lens L1 and coupled into a SM fiber via
lens L2. The filter removes the pump light and allows only the
downconverted light to be coupled into the fiber.
PPSLTSource-f2.eps.}
\end{figure}

\begin{figure}[htbp]
 \centering \includegraphics[width=6in]{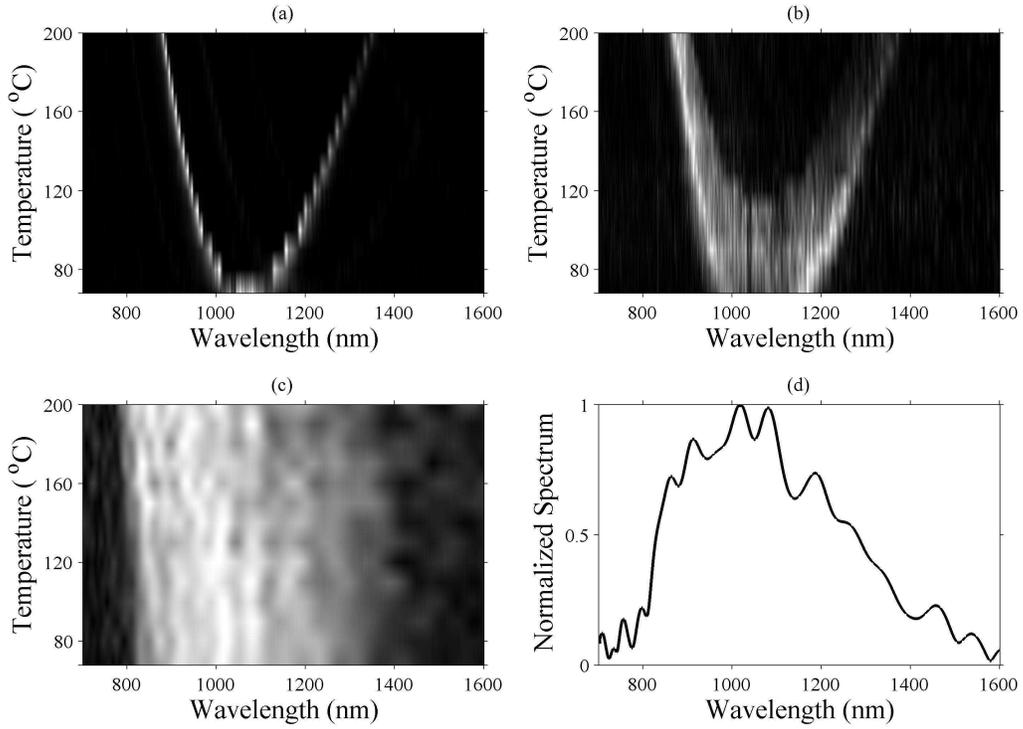}
\caption{Brightness images, in which brightness is proportional to
the measured power spectral density of the emission, at various
temperatures, from: (a) the unchirped structure; (b) the
medium-chirped structure; and (c) the maximum-chirped structure. (d)
Measured normalized spectrum for light from the maximum-chirped
structure at a temperature of 80$^\circ$ C. The results bear
considerable resemblance to the calculations displayed in Fig. 1.
ExperimentalSpectra-f3.eps.}
\end{figure}

\begin{figure}[htbp]
 \centering \includegraphics[width=6in]{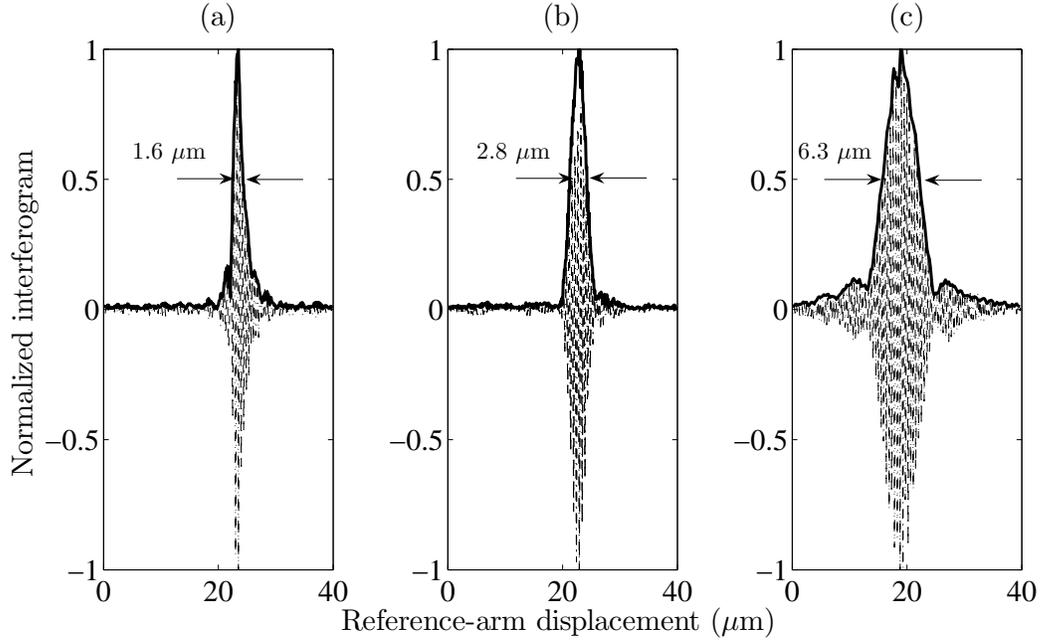}
\caption{Normalized interferograms and their envelopes vs.
reference-arm displacement for a mirror sample (A-scans). In all
cases, the step size used in constructing the interferograms was 100
nm and the duration of the counting-time windows was 300 ms. (a)
Downconversion/superconducting-detector (SPDC/SSPD). The FWHM of the
interferogram envelope was 1.6 $\mu$m. The highest resolution was
achieved with this combination. (b)
Downconversion/avalanche-detector (SPDC/SPAD). The FWHM of the
interferogram envelope was 2.8 $\mu$m. (c)
Superluminescence/superconducting-detector (SLD/SSPD). The FWHM of
the interferogram envelope was 6.3 $\mu$m. Resolution-f4.eps.}
\end{figure}

\begin{figure}[htbp]
 \centering \includegraphics[width=6in]{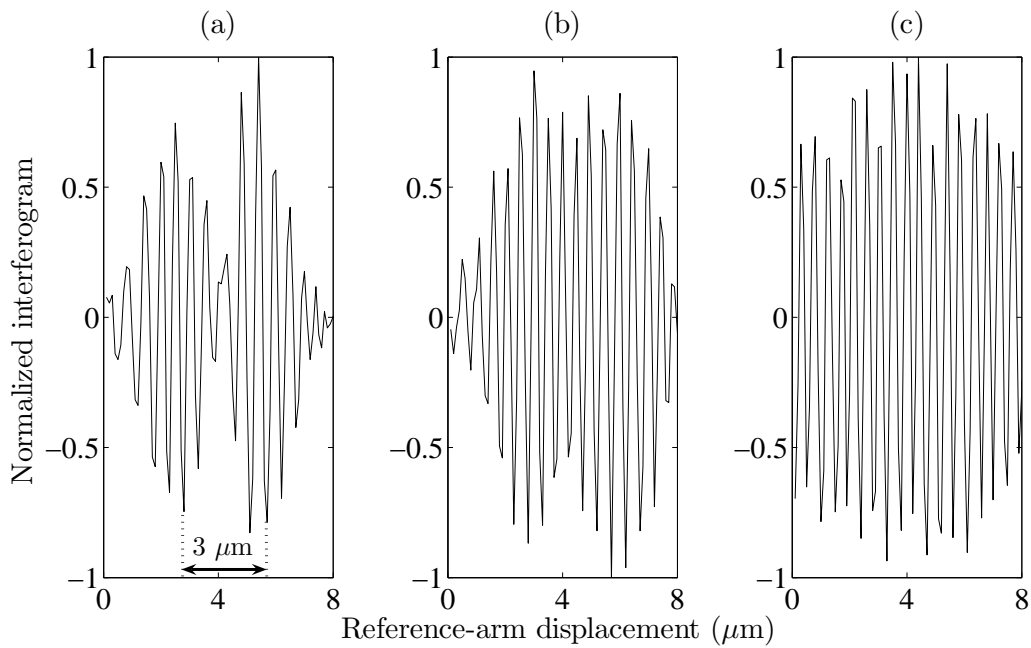}
\caption{Normalized interferograms vs. reference-arm displacement
for a pellicle sample (A-scans). In all cases, the step size used in
constructing the interferograms was 100 nm and the duration of the
counting-time windows was 300 ms. (a)
Downconversion/superconducting-detector (SPDC/SSPD). This
combination permitted reflections from the two surfaces to be
resolved. (b) Downconversion/avalanche-detector (SPDC/SPAD). The two
surfaces were not resolved. (c)
Superluminescence/superconducting-detector (SLD/SSPD). The two
surfaces were not resolved. Pellicle-f5.eps.}
\end{figure}

\begin{figure}[htbp]
 \centering \includegraphics[width=6in]{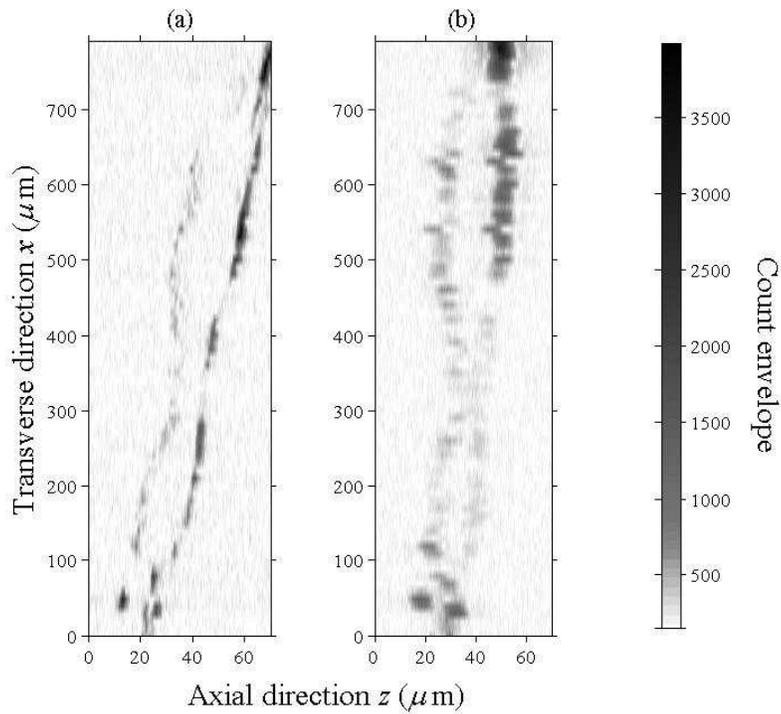}
\caption{Two-dimensional ($xz$) B-scans of an onion-skin sample. (a)
Scan collected using broadband downconversion light and a
superconducting detector (SPDC/SSPD). (b) Scan collected from the
same onion-skin sample using superluminescence light and the same
superconducting detector (SLD/SSPD). Higher axial resolution is
attained by using downconversion, by virtue of its broader
bandwidth. These cross-sectional views of the tissue highlight the
relatively large reflectances at cellular surfaces, which stem from
refractive-index discontinuities. Onion-f6.eps.}
\end{figure}

\end{document}